\documentclass[%
 reprint,
 superscriptaddress,
nofootinbib,
 amsmath,amssymb,
 aps,
 pra,
floatfix,
]{revtex4-2}

\usepackage{graphicx}
\usepackage{dcolumn}
\usepackage{bm}
\usepackage{siunitx}
\usepackage{hyperref}
\usepackage[capitalise]{cleveref}
\usepackage{physics}
\usepackage{booktabs}
\usepackage{dsfont}
\usepackage[acronym]{glossaries}
\usepackage{comment}
\usepackage[normalem]{ulem}
\newacronym{qm}{QM}{Quantum Mechanics}
\newacronym{qkd}{QKD}{Quantum Key Distribution}
\newacronym{locc}{LOCC}{Local Operations and Classical Communication}
\newacronym{cptp}{CPTP}{Completely Positive Trace Preserving}
\newacronym{qec}{QEC}{Quantum Error Correction}
\newacronym{nisq}{NISQ}{Noisy Intermediate-Scale Quantum}
\newacronym{epl}{EPL}{Extreme Photon Loss}
\newacronym{sc}{SC}{Single-Click}
\newacronym{dc}{DC}{Double-Click}
\newacronym{ga}{GA}{Genetic Algorithm}
\newacronym{povm}{POVM}{Positive Operator Valued Measure}
\newacronym{swapasap}{SWAP-ASAP}{Network protocol that performs entanglement SWAP As Soon As Possible}

\glsdisablehyper  

\renewcommand{\vec}[1]{\bm{#1}}

\newcommand{\Id}{\mathds{I}}
\newcommand{\vx}{\vec{x}}
\newcommand{\vy}{\vec{y}}

\begin{document}

\preprint{APS/123-QED}

\title{Reducing hardware requirements for entanglement distribution via joint hardware-protocol optimization}

\author{Adrià Labay Mora}
\email{alabay@ifisc.uib-csic.es}
\affiliation{%
 Institute for Cross Disciplinary Physics and Complex Systems (IFISC) UIB-CSIC, Campus Universitat Illes Balears, Palma de Mallorca, Spain.
}
\author{Francisco Ferreira da Silva}%
\author{Stephanie Wehner}
\affiliation{%
 QuTech, Delft University of Technology, Lorentzweg 1, 2628 CJ Delft, The Netherlands
}%
\affiliation{Quantum Computer Science, EEMCS, Delft University of Technology, Lorentzweg 1, 2628 CJ Delft, The Netherlands}
\affiliation{Kavli Institute of Nanoscience, Delft University of Technology, Lorentzweg 1, 2628 CJ Delft, The Netherlands}

\date{\today}

\begin{abstract}
We conduct a numerical investigation of fiber-based entanglement distribution over distances of up to $\SI{1600}{km}$ using a chain of processing-node quantum repeaters.
We determine minimal hardware requirements while simultaneously optimizing over protocols for entanglement generation and entanglement purification, as well as over strategies for entanglement swapping.
Notably, we discover that through an adequate choice of protocols the hardware improvement cost scales linearly with the distance covered.
Our results highlight the crucial role of good protocol choices in significantly reducing hardware requirements, such as employing purification to meet high-fidelity targets and adopting a SWAP-ASAP policy for faster rates.
To carry out this analysis, we employ an extensive simulation framework implemented with NetSquid, a discrete-event-based quantum-network simulator, and a genetic-algorithm-based optimization methodology to determine minimal hardware requirements.
\end{abstract}

\maketitle

\section{Introduction}

Major efforts are underway towards the development of the quantum internet \cite{Wehnereaam9288,kimble2008quantum} which can in principle enable information-theoretically secure communication through quantum key distribution \cite{shor2000simple,bennett2014qkd}. Other applications include secure access to remote quantum computers~\cite{broadbent2009universal}, more accurate clock synchronization~\cite{jozsa2000quantum}, scientific applications such as combining light from distant telescopes to improve observations \citep{sidhu2021satellite}, and distributed quantum computation \citep{cirac1999distributed}.

Most aforementioned applications require the generation of entangled states between the quantum devices executing them~\cite{Wehnereaam9288}. Entanglement generation has thus far only been demonstrated experimentally at up to metropolitan distances in fiber~\cite{hensen_loopholefree_2015,yu2020entanglement}, and scaling up is challenging due to photon loss in optical fiber which increases exponentially with distance~\cite{ekert2000physics}. Classically, direct amplification is used to overcome photon loss but this is forbidden in the quantum case by the no-cloning theorem \cite{wootters1982nocloning}. 

A possible solution is the use of intermediate nodes, known as quantum repeaters~\cite{briegel1998optimising,munro2015inside}. These devices can in theory extend the distance over which entangled states can be faithfully transmitted. However, the quality of the entanglement transmitted using quantum repeaters decays exponentially with the number of repeaters used~\cite{briegel1998optimising}. Despite recent progress on both the hardware and theory fronts, a scalable quantum repeater has yet to be demonstrated~\cite{langenfeld2021quantum,bhaskar2020experimental,hermans2022qubit}. 

Part of the challenge in developing quantum repeaters is that hardware requirements are not fully known. There have been many investigations of such requirements (see~\cite{avis2022requirements} and references therein). These requirements depend on multiple factors, such as which protocols are employed to generate and distribute entanglement, the placement of the repeaters (which will likely be constrained by existing fiber infrastructure~\cite{da2023requirements,rabbie2022designing}), and how many repeaters are used. To the best of our knowledge, there has yet to be a study of hardware requirements that simultaneously investigates the protocols executed by the nodes in the repeater chain and the nodes' hardware in the presence of time-dependent noise. Our study aims to fill this gap.

Different protocol choices likely result in different hardware requirements, as they put emphasis on different hardware properties. For example, entanglement-purification protocols can be used to enhance the quality of shared links but necessitate higher-quality gates, as they require that more operations are performed. Many such questions arise when one considers all the building blocks required to generate entanglement over long distances using quantum repeaters.

In this work, we investigate minimal hardware requirements for quantum-repeater chains spanning up to $\SI{1600}{km}$.
We consider requirements for achieving (a) a fidelity of $0.8$ and a rate of $\SI{1}{Hz}$ and (b) a fidelity of $0.9$ and a rate of $\SI{0.1}{Hz}$. 
We remark that a secret key can be distilled from states satisfying each of these fidelity targets using the BB84 quantum key distribution protocol~\cite{bennett2014qkd} in its entanglement-based version~\cite{bennett1992quantum} (with two-way communication being required for states fulfilling only the lower-fidelity target~\cite{gottesman2003proof}).
This is shown in \cref{appendix:fidelity_qkd}.
We expect that by picking these two targets we will be able to probe two different parameter regimes in terms of which protocols perform best and what are the corresponding hardware requirements.
We determine minimal hardware requirements while optimizing over the entanglement generation protocol used for establishing nearest-neighbor links, the purification protocol, and the global network protocol that controls the sequence of actions in the chain \cite{briegel1998optimising,coopmans_netsquid_2020}.
We combine a simulation-based approach that allows us to accurately account for the effects of time-dependent noise with an optimization methodology based on \glspl{ga} to determine minimal hardware requirements \cite{da_silva_optimizing_2020,chehimi2023scaling}.

We find that meeting the performance targets we set at distances of over $\SI{200}{km}$ is only possible using quantum repeaters, with a spacing of roughly $\SI{100}{km}$ performing best.
Further, we find that an adequate choice of protocols minimizes the required hardware improvement over the experimental state-of-the-art.

\section{Methodology}
\label{sec:methodology}

\begin{figure*}
    \centering
    \includegraphics[width=\linewidth]{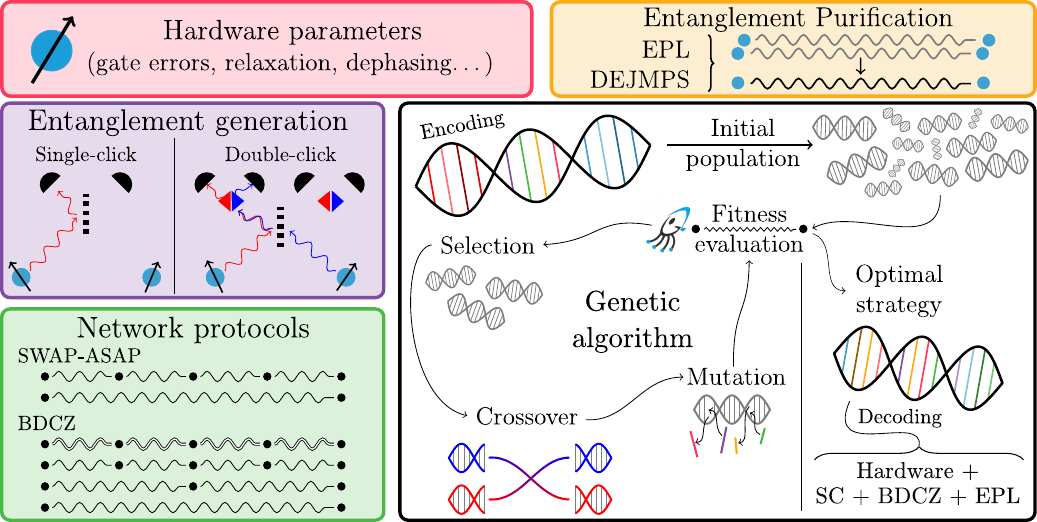}
    \caption{Building blocks of the repeater chain and optimization method we consider. The chain consists of $2^n+1$ ($n\in \mathds{N}$) identical quantum devices with $2$ end nodes and $2^n - 1$ repeaters which are equally spaced in a line with total distance $d$. Each node hardware is parameterized in terms of noise parameters as detailed in Section~\ref{sec:hardware_model}. Entangled states can be generated between neighboring nodes using single-click or double-click, both heralded entanglement generation protocols. These links can be purified using one of the two studied protocols: EPL or DEJMPS. Both consume two elementary links to probabilistically yield a higher fidelity one. Finally, the network protocols orchestrate the global sequence of actions in the chain. We distinguish between SWAP-ASAP (no entanglement purification) and BDCZ (a nesting level strategy with purification). The network is simulated using NetSquid from which we extract the fidelity and rate of the entangled states shared between the two end nodes. The genetic algorithm searches in the space of all hardware and protocol parameters for the solution with the lowest improvement over state-of-the-art parameters which still satisfy the target values. }
    \label{fig:pp_repchain}
\end{figure*}

In this section, we introduce our approach to finding minimal requirements for quantum-repeater hardware. We elaborate on how we model hardware, the repeater protocols we consider, and the optimization methodology used. A visual summary of the contents of this section can be seen in \cref{fig:pp_repchain}.
We note that similar methodologies have been employed earlier~\cite{da_silva_optimizing_2020,avis2022requirements,da2023requirements}.

\subsection{Hardware model}
\label{sec:hardware_model}
Multiple different physical systems are being investigated as possible hardware platforms for quantum repeaters. These include color centers in diamond~\cite{ruf2021quantum}, trapped ions~\cite{duan2010colloquium, reiserer2015a}, neutral atoms~\cite{reiserer2015a,langenfeld2021quantum} and quantum dots~\cite{gao2012observation,li2016heralded}. Despite impressive recent developments~\cite{pompili2021realization, bhaskar2020experimental} a scalable quantum repeater has yet to be demonstrated. Modeling of quantum-repeater hardware can be useful in understanding how hardware limitations impact the repeater's performance, shedding light on, for example, which hardware parameters require the most improvement.

We consider a simplified platform-agnostic model for quantum nodes that aims to capture the most relevant noise sources common to all processing-node repeaters, i.e., repeaters that can not only store quantum information but also perform quantum gates.
This renders our results relevant to all of them.
We assume that nodes have a quantum memory of $N_{qb}$ fully-connected qubits (i.e., two-qubit gates can be executed between any two qubits in the memory) which decohere with characteristic relaxation time $T_1$ and dephasing time $T_2$. This means that if a state $\rho$ is stored in memory, it undergoes amplitude damping:
\begin{equation}
\rho \rightarrow E_0 \rho E_0^{\dagger} +  E_1 \rho E_1^{\dagger}\ ,
\end{equation}
where the Kraus operators $E_0$ and $E_1$ are given by
\begin{equation}
    E_0 = \ket{0}\bra{0} + \sqrt{1-p_{T_1}}\ket{1}\bra{1}, E_1 = \sqrt{p_{T_1}}\ket{0}\bra{1}\ ,
\end{equation}
and the probability $p_{T_1}$ for the amplitude damping process is given by:
\begin{equation} \label{eq:pp_t1}
    p_{T_1} = 1 - e^{-t/T_1}\ .
\end{equation}
States stored in memory also undergo dephasing:
\begin{equation}
    \rho \rightarrow (1-p_{T_2})\rho + p_{T_2} Z\rho Z\ ,
\end{equation}
with the probability $p_{T_2}$ for the dephasing process being given by
\begin{equation} \label{eq:pp_t2}
    p_{T_2} = \frac{1}{2}\left(1 - e^{t/T_2} e^{t/2T_1}\right)\ .
\end{equation}
All qubits are assumed to have the same $T_1$ and $T_2$.

Arbitrary single-qubit rotations can be performed on every qubit and are subject to depolarizing noise with probability $p_1$.
Similarly, two-qubit gates are subject to depolarizing noise with probability $p_2$.
We model gate noise by first applying the operation perfectly and then applying the noise channel.
The $d$-dimensional depolarizing noise channel acts on a state $\rho$ as follows,
\begin{equation}
    \rho \rightarrow p \rho + (1-p)\frac{\Id}{d},
\end{equation}
where $\Id$ is the $d$-dimensional identity matrix. The single-qubit (two-qubit) case then corresponds to $d=2$ ($d=4$). All qubits can be measured in the $Z$ basis, with a bit-flip error probability $\xi_0$ and $\xi_1$ of obtaining the wrong outcome. We assume that repeaters can only attempt to generate entanglement with one neighbor at a time, as is the case for many proposed quantum-repeater platforms~\cite{ruf2021quantum,avis2022requirements}.

\subsection{Protocols for end-to-end entanglement generation}
\label{sec:protocols}
We aim to generate entanglement between two distant end nodes of the type described in \cref{sec:hardware_model} which are connected by a chain of quantum repeaters realized with the same type of nodes. This task is broken down into the generation of entanglement between neighboring nodes, the connection of two of these short links into a longer one, and link purification. In this section, we elaborate on protocols for each of these tasks.

\subsubsection{Nearest-neighbor entanglement generation} \label{sec:nneg}
We consider heralded entanglement generation protocols~\cite{azuma2022quantum}.
These rely on the existence of a heralding station which we assume to be placed equidistantly between two neighboring nodes. The station consists of a protocol-dependent combination of beam splitters and photon detectors. 
In particular, we investigate single~\cite{cabrillo1999creation} and double-click~\cite{barrett2005efficient} entanglement-generation protocols, whose success is heralded by the detection of one and two photons, respectively.
In both protocols, entanglement generation is done in successive attempts, with the participating nodes learning if they have been successful in generating entanglement at the end of each attempt.
We assume that each entanglement generation attempt takes time $L/c$ where $L$ is the fiber distance between the two nodes and $c \sim \SI{200000}{km.s^{-1}}$ is approximately the speed of light in fiber.
The probability $p_{\text{det}}$ of a photon emitted by a node being successfully detected at the midpoint station is given by
\begin{equation}
    p_{\text{det}} = p_{\text{emd}} \times 10^{-(\alpha_{\text{att}}/10)(L/2)},
\end{equation}
where $\alpha_{\text{att}} = \SI{0.2}{dB.km^{-1}}$ is the fiber's attenuation coefficient and $p_{\text{emd}}$ is the probability that an emitted photon is detected, given that it was not lost in fiber.
This parameter then combines multiple loss sources, such as the probability of emitting the photon in the right mode, the probability of collecting it into the fiber, and the probability of detecting it given that it arrived at the detector.

To first approximation, the entangled states $\rho_{\text{sc}}$ generated with the \gls{sc} protocol are of the type~\cite{humphreys_deterministic_2018,rozpedek_optimizing_2018}
\begin{equation}
    \rho_{\text{sc}} = (1-\alpha) \ket{\psi^{\pm}}\bra{\psi^{\pm}} + \alpha \ket{11}\bra{11},
\label{eq:single_click_state}
\end{equation}
where $\alpha$ is a tunable parameter and $\ket{\psi^{\pm}} = (\ket{01} \pm \ket{10})/\sqrt{2}$ is the ideal Bell state.
The fidelity of this state to the ideal Bell state is $F=1-\alpha$, and the probability of successfully generating entanglement with the single-click protocol is given by $2p_{\text{det}}\alpha$.
This has two important consequences: (i) in this protocol, fidelity can be traded off against success probability (and hence entanglement generation rate) and (ii) states generated with this protocol are imperfect (i.e., they have non-unit fidelity to the ideal Bell state) even under the assumption of perfect hardware. 

We consider multiple noise sources that reduce the fidelity to the state in~\cref{eq:single_click_state}.
Specifically, the probability of emitting two photons instead of one $p_{\mathrm{double}}$, the phase uncertainty $\sigma_\phi$ acquired while traveling through the fiber~\cite{kalb2017entanglement} and the Hong-Ou-Mandel visibility $V$ of the interfering photons~\cite{hong1987measurement}.
The visibility is defined as $V = 1 - C_{min}/C_{max}$~\cite{bouchard2020two}, where $C_{min}$ is the probability that two photons being interfered at a 50:50 beamsplitter are detected at two different detectors when indistinguishability is optimized and $C_{max}$ is the same probability when photons are made distinguishable (e.g., by having them arrive at different times).

To account for these effects we define the state efficiency
\begin{equation}
    \eta_f = \frac{1 + \sqrt{V}}{2}(1 - p_{ph})\ ,
\end{equation}
where $p_{ph}$ can be derived from the hardware parameters $p_{\mathrm{double}}$ and $p_\phi = [1 - \exp(-\sigma_\phi^2/2)]/2$ as
\begin{equation}
\small
    p_{ph} = (1 - p_\phi)p_d(1-p_{\mathrm{double}}) + p_\phi[p_{\mathrm{double}}^2 (1-p_{\mathrm{double}})^2]\ .
\end{equation}
Hence, the fidelity of the elementary link generated using the single-click protocol is 
\begin{equation} \label{eq:paper_fsc}
    f_{SC} = (1 - \alpha) \eta_f \ .
\end{equation}

States generated with the \gls{dc} protocol are, under the assumption of perfect hardware, ideal Bell states. Therefore, in this protocol, there is no inherent limitation on the achievable fidelity.
We do however model two noise sources, namely the Hong-Ou-Mandel visibility $V$ and the light-matter-interface fidelity $f_{lm}$, i.e., we allow for the possibility of depolarizing noise in the light-matter state generated at the nodes.
The state generated then looks like this:
\begin{equation}
\begin{split}
    \rho_{\mathrm{DC}} =& \frac{f_{lm}}{2} [(1\pm V)\op{\Phi_{01}} + (1 \mp V) \op{\Phi_{11}}]\\ &+ \frac{1-f_{lm}}{2} [\op{00} + \op{11}].
\end{split}
\end{equation}
Its fidelity is $f_{DC} = f_{lm} (1 + V) / 2$.
The double-click protocol succeeds with probability $1/2 p_{\text{det}}^2$.
The scaling of the success probability with $p_{\text{det}}$, and consequently with the fiber length, is thus less favorable for the double-click protocol when compared with the single-click protocol.


\subsubsection{Entanglement swapping}
Entanglement swapping is a protocol based on quantum teleportation~\cite{bennet1895teleportation} that effectively generates longer entangled links by consuming shorter ones. Imperfections in the links and the gates used in the swap circuit cause the fidelity to decay exponentially with the number of swaps \cite{briegel1998optimising}.
We assume the entanglement swap circuit is implemented through a Hadamard gate, a CNOT gate, and measurements in the computational basis. As discussed in \cref{sec:hardware_model}, the gates suffer from depolarizing noise, and the measurements from bit-flip errors.

\subsubsection{Entanglement purification}
Entanglement purification protocols probabilistically generate fewer higher-fidelity entangled pairs from many lower-fidelity ones. We focus in particular on DEJMPS~\cite{deutsch1996distillation} and \gls{epl}~\cite{dam_multiplexed_2017, kalb2017entanglement}, which are both two-to-one protocols. 

Concretely, the EPL protocol consists of applying CNOT gates between the entangled pairs (using one of the pairs as controls and the other pair as targets), measuring the target qubits, and keeping the entangled pair corresponding to the control qubits if the outcomes are both $1$ in the $Z$ basis. For any other combination of measurement outcomes, the remaining entangled pair is discarded. This protocol yields maximally entangled states when applied to states of the form $\rho_{\text{sc}}$ (see \cref{eq:single_click_state}), succeeding with probability $\frac{1}{2}(1-\alpha)^2$. In fact, EPL has been shown to be optimal for such states, in the sense that (i) no other purification protocol achieves a higher fidelity, and (ii) no other protocol achieves the same fidelity with higher success probability~\cite{rozpedek_optimizing_2018}.

The DEJMPS protocol starts with Alice and Bob applying the unitaries $U_A$ and $U_B$, respectively, to each of their qubits.
$U_A$ is defined as
\begin{equation}
    \ket{0} \rightarrow \frac{1}{\sqrt{2}}(\ket{0} - i\ket{1}),\quad \ket{1} \rightarrow \frac{1}{\sqrt{2}}(\ket{1} - i\ket{0}),
\end{equation}
and $U_B$ is defined as
\begin{equation}
    \ket{0} \rightarrow \frac{1}{\sqrt{2}}(\ket{0} + i\ket{1}),\quad  \ket{1} \rightarrow \frac{1}{\sqrt{2}}(\ket{1} + i\ket{0}).
\end{equation}
They then apply CNOTs and perform measurements in the $Z$ basis just as in the EPL protocol, but in this case accept if both measurement outcomes are equal (i.e., also in the $00$ case).
The output fidelity of the protocol for an input state $\rho$ is
\begin{equation}
    F = \frac{A^2 + B^2}{p_{\text{succ}}},
\end{equation}
with a probability of success
\begin{equation}
    p_{\text{succ}} = (A + B)^2 + (C + D)^2,
\end{equation}
where
\begin{align*}
    &A = \bra{\Phi_+}\rho\ket{\Phi_+},
    &B = \bra{\Phi_-}\rho\ket{\Phi_-}, \\ 
    &C = \bra{\Psi_+}\rho\ket{\Psi_+},
    &D = \bra{\Psi_-}\rho\ket{\Psi_-},
\end{align*}
and $\ket{\Phi_{\pm}}=\frac{1}{\sqrt{2}}(\ket{00} \pm \ket{11})$ and $\ket{\Psi_{\pm}} = (\Id \otimes X)\ket{\Phi_\pm}$.
DEJMPS has been shown to be optimal for Bell-diagonal states of rank up to $3$~\cite{rozpedek_optimizing_2018}, for the same definition of optimality given for EPL concerning states of the form $\rho_{\text{sc}}$.

\subsubsection{Repeater chain protocols}
\begin{figure}
    \centering
    \includegraphics[width=\linewidth]{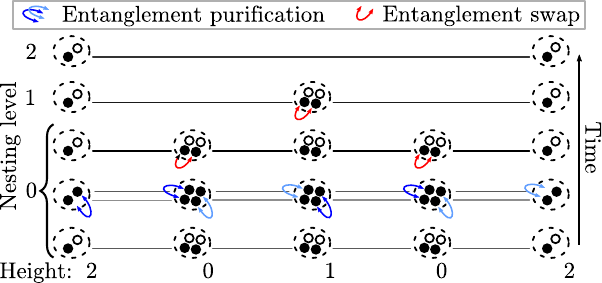}
    \caption{Diagram of the BDCZ protocol on a 5-node network. The height of each node determines the nesting level where entanglement swap is performed. Qubits at the nodes are represented as circles which are filled if they store an entangled state. The network sequentially creates two pairs of entangled states between neighboring quantum nodes. The fidelity of such links is small (depicted with the color intensity, lower fidelity corresponds to grayer colors) and suffers from time dephasing while in memory (\cref{eq:pp_t1,eq:pp_t2}). Two links shared between the same pair of nodes are purified, leading to a high-fidelity state that can be swapped. Afterwards, the nesting level increases and the nodes with lower height are no longer needed in the process. This process is repeated until end-to-end entanglement has been established.}
    \label{fig:bdcz_paper}
\end{figure}

Repeater chain protocols orchestrate the subprotocols described above in order to generate end-to-end entanglement across a repeater chain. We investigate two such protocols, SWAP-ASAP~\cite{inesta2022optimal,coopmans_netsquid_2020} and BDCZ~\cite{briegel1998optimising}. The first one consists of a swap-as-soon-as-possible strategy, in which repeater nodes perform an entanglement swap whenever they hold two entangled pairs. The second one is a nested strategy that combines entanglement generation and entanglement purification. In this protocol, nodes are assigned a height that depends on their relative position in the chain and determines when they are allowed to perform an entanglement swap (see \cref{fig:bdcz_paper}). Whenever two nodes of the same height are connected by an entangled pair they can either decide they are ready to swap or generate more entangled pairs and then perform purification. Swapping results in nodes of larger height being connected, for which purification is a high-stake endeavor, as failure implies regenerating entangled pairs at the initial height.

Both the hardware modeling introduced in \cref{sec:hardware_model} and the protocols introduced in this section are simulated using the discrete-event-based quantum-network-simulator NetSquid, building on previous work done in~\cite{coopmans_netsquid_2020, dasilva2021optimizing}. These simulations allow us to evaluate the performance achieved by different sets of hardware parameters.

\subsection{Optimization algorithm} \label{sec:optimization}

In Ref.~\cite{dasilva2021optimizing}, the authors restate the question of finding minimal hardware requirements as an optimization problem. 
They define minimal hardware requirements as the hardware parameters requiring the smallest improvement over state-of-the-art parameters that fulfill given performance targets.
This is quantified by mapping sets of hardware parameters to an improvement cost via a cost function that can then be minimized.
Here, we expand on the work done in Ref.~\cite{dasilva2021optimizing} by also optimizing over protocol parameters.
A common theme among the protocols discussed in \cref{sec:protocols} is rate-fidelity trade-offs.
For example, employing a single-click instead of a double-click entanglement generation protocol will typically result in higher entanglement generation rates, at the expense of lower fidelities.
It should then be expected that different protocol choices result in different minimal hardware requirements, and therefore that optimizing hardware and protocol parameters simultaneously should lead to less stringent hardware requirements.

The cost function that encodes the question we aim to answer is
\begin{equation} \label{eq:cost_function}
    T_C(\vec{x}; \vec{y}) = \mathcal{C}(\vec{x}) + A \mathcal{P}(\vec{y})\ ,
\end{equation}
which contains two terms. The first term,
\begin{equation}
    \mathcal{C}(\vec{x}) = \sum_j [\log_{x_{\mathrm{base}}^j}(x^j)]^{-1}\ ,
\end{equation}
is the cost of improving a set of hardware parameters from a given baseline $\vec{x}_{\mathrm{base}}$ to $\vec{x}$. This is the \textit{hardware cost}. Here, $\vec{x} = (x_1,x_2,\dots)$ is a vector containing the value of the optimization parameters.
The second term,
\begin{equation}
    \mathcal{P}(\vy) = \sum_k [1 + (y_{\mathrm{target}}^k - y^k)^2] \Theta(y_{\mathrm{target}}^k - y^k)\ ,
\end{equation}
is the penalty assigned for not meeting the performance targets $\vy_{\mathrm{target}}$. It vanishes if all targets are met and is larger as the gap between the targets and the achieved performance grows. Finally, the hyperparameter $A$ is chosen such that $C(\vx) < A P(\vy)$ for any allowed parameter set $\vx$, effectively ensuring that the performance targets are hard constraints.
The goal of the optimization procedure is to find the set of parameters that minimize the hardware cost while satisfying the performance targets.
This set of parameters is the minimal hardware requirements.

To find minimal hardware requirements, we must solve the optimization problem we just defined. We do so by employing a genetic-algorithm-based optimization methodology. These algorithms have advantages over deterministic methods like gradient descent when little is known about the landscape of the function to be optimized or if it is non-differentiable~\cite{goldberg1989genetic}. Moreover, local methods tend to converge to the local minimum closest to the starting point, thus they often fail to find the global optimum in problems with multiple minima. \glspl{ga} can avoid this through a strategy combining exploration and exploitation. The approach we employ here is based on the one introduced in Ref.~\cite{dasilva2021optimizing} which is summarized in \cref{fig:pp_repchain}. A collection of 120 random optimization parameters is generated $\{ \vx_a \}$ defining the initial population. Then, the cost function \eqref{eq:cost_function} is evaluated for each parameter set $\vx_a$, where the target values $\vy_a = (F, R)$ are evaluated from an average over 100 realizations of the Netsquid repeater-chain simulation.
Averaging was done due to the stochastic nature of the simulations.
We empirically found that 100 realizations struck a good balance between accuracy and computation time.
Parameter sets achieving low cost are selected for the next generation of the \gls{ga} together with new sets of parameters created using crossover and mutation genetic processes.
In the final round, after 500 iterations, the parameter set $\vx_{\mathrm{min}}$ with minimum cost function is selected as the optimal combination of hardware and protocol parameters with the lowest hardware improvement over state-of-the-art hardware parameters.
This a standard way to terminate an evolutionary algorithm~\cite{jain2001termination}.

To the previous procedure, we add an extra step which consists of an iterative local search optimization over the best solution found by the \gls{ga} to ensure exploitation of the minimum found \cite{johnson1988easy}. This deterministic optimization algorithm is also a gradient-free method that with few function evaluations can, if possible, reduce the total parameter cost. We note that it only searches for possible reduced minima on a region around the optimal hardware parameters without changing the protocols. That is, we assume the \gls{ga} has found the combination of protocols leading to the lowest hardware cost, but has not been able to reach it. 

The complete optimization procedure -- including 500 iterations of the \gls{ga} with 120 individuals per generation and 100 repetitions of the network simulation per individual -- takes about $\SI{30}{min}$ and $\SI{3}{days}$ for two- and nine-node networks on a high-performance supercomputer \cite{cartesius}. The code used for the simulations is open-source and can be found in Ref.~\cite{2023sourcecode}. The data extracted with the best individuals for each network scenario can also be found in Ref.~\cite{2023replicationdata}.

For more details on the optimization methodology, see \cref{apx:optimization}.
\subsection{State-of-the-art parameters}
Determining minimal hardware requirements as described in \cref{sec:optimization} is done with respect to a baseline. We used experimentally realized parameters in color center experiments to determine these parameters. This choice was made as color centers have been used to demonstrate several quantum-networking primitives, such as remote entanglement generation~\cite{humphreys_deterministic_2018}, long-lived quantum memories~\cite{bradley2019ten}, entanglement purification~\cite{kalb2017entanglement} and entanglement swapping in a three-node network~\cite{hermans2022qubit}.
The parameter values used are shown in \cref{tab:baseline_parameters}. Details on how they were determined can be found in~\cref{apx:validation}.
\begin{table}
    \centering
    \caption{State-of-the-art color-center-based hardware parameters involved in the optimization. All gate-based errors ($p_1$, $p_2$, $\xi_{0/1}$ and $p_{init}$) are improved simultaneously with the same cost. The last two parameters $\eta_f$ and $f_{elem}$ apply only to single-click (SC) and double-click (DC) respectively as explained in \cref{sec:nneg}. The value for the parameters has been obtained from state-of-the-art color-center experiments as explained in Ref.~\citep{coopmans_netsquid_2020}.}
    \label{tab:baseline_parameters}
    \begin{tabular}{@{}l|c@{}}
        \toprule
        \textbf{Parameter} & \textbf{Value} \\ \midrule
        $p_{1}$ (Single-qubit gate error) & $(4/3)0.001\%$ \\
        $p_{2}$ (Two-qubit gate error) & $0.02\%$ \\
        $\xi_0,\ \xi_1$ (Readout error) & $0.05\%,\ 0.005\%$\\
        $p_{init}$ (Initialisation error) & $0.02\%$ \\
        $T_1$ (Relaxation time) & $\SI{1}{h}$ \\
        $T_2$ (Dephasing time) & $\SI{1}{s}$ \\
        $p_{\text{emd}}$ (Photonic efficiency excluding fiber) & $0.46\%$ \\
        $\eta_f$ (SC) (State efficiency) & $91.96\%$ \\
        $f_{elem}$ (DC) (Elementary link fidelity) & $92\%$ \\ \bottomrule
    \end{tabular}
    \end{table}
We then optimize over all hardware parameters shown in~\cref{tab:baseline_parameters} as well as over entanglement generation, whether SWAP-ASAP or BDCZ is used, which purification protocol is employed, and how many times purification is done. In case single-click is employed we optimize over the bright-state parameter $\alpha$ as well.

\section{Results} 
We choose fidelity $F_t$ and entanglement generation rate $R_t$ as our performance metrics.
Concretely, we pick two pairs of target values: (a) $F_t = 0.8$ and $R_t = \SI{1}{Hz}$ and (b) $F_t = 0.9$ and $R_t = \SI{0.1}{Hz}$. As discussed in \cref{sec:methodology}, any choice of protocols for entanglement generation and purification implies trade-offs. For example, the success probability of \gls{sc} scales more favorably compared to \gls{dc}, but this comes at the expense of inherently lower fidelity. Therefore, we expect that by picking different targets we will be able to probe different regimes in terms of which protocols and hardware improvements are found to be optimal.

Hence, in each simulation for a particular distance and number of repeaters, the genetic algorithm explores the full space of hardware parameters and protocols introduced in \cref{sec:hardware_model} and \cref{sec:protocols} respectively. We here assume that the number of quantum repeaters used is a quantity to be optimized and to which no cost is assigned. As the BDCZ protocol is only well defined for numbers of nodes $N = 2^n-1$ due to its hierarchical swap structure, we restrict ourselves to configurations verifying this condition.
\subsection{Optimal hardware cost}
\label{sec:cost_scaling}
\begin{figure}
    \centering
    \includegraphics[width=\linewidth]{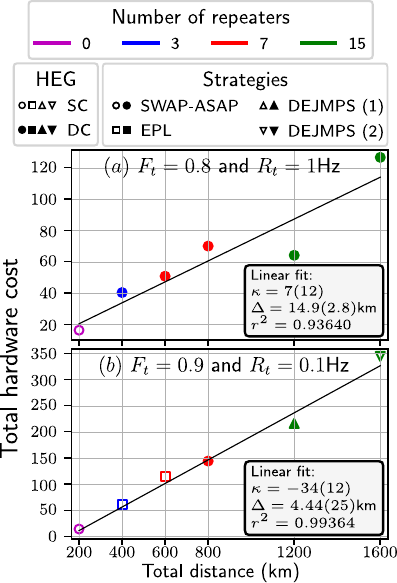}
    \caption{Total hardware cost (see \cref{eq:cost_function}) resulting from the best combination of hardware parameters, network protocols, and number of repeaters for a specific total distance, as found by our method. (a) The target values are $F_t = 0.8$ and $R_t = \SI{1}{Hz}$ and (b) $F_t = 0.9$ and $R_t = \SI{0.1}{Hz}$. In both panels, the optimal protocol strategies are denoted by the color, shape, and filling of the markers. The color coding indicates the number of repeaters used ($N_{nodes} = N_{qr} + 2$), while the shape represents the type of purification and network protocol applied (circles for SWAP-ASAP, squares for BDCZ with EPL,  up-triangles for BDCZ with one round of DEJMPS and down-triangles for BDCZ with two rounds of DEJMPS in the first nesting level). Filled and unfilled markers indicate the use of DC and SC, respectively. The black line represents a linear fit of the cost values with the fit parameters shown in the inset.}
    \label{fig:paper_everything}
\end{figure}

We show in \cref{fig:paper_everything} the cost, as defined in \cref{eq:cost_function}, of distributing entanglement satisfying targets (a) and (b) over distances ranging from $\SI{200}{km}$ to $\SI{1600}{km}$. This corresponds to the best solutions found by the GA after the local iterative minimization of the hardware parameters. 

We initially observe that the total hardware cost follows a linear scaling behavior in relation to the covered distance, as depicted by the linear fits depicted in the figure. This finding is unexpected, considering that multiple quantities associated with quantum repeater chains typically exhibit exponential scaling with respect to the distance or the number of repeaters employed. It is the case for, among others, the photon transmission probability with distance and the fidelity decay with the number of repeaters~\cite{briegel1998optimising}. We must however note that the cost function we employed does not explicitly assign a cost to placing quantum repeaters. It does so implicitly, as using different numbers of repeaters incurs different hardware requirements and hence different costs. Nevertheless, this provides evidence that optimizing the choice of protocols and the number of repeaters employed allows for more favorable scaling of hardware requirements. This highlights the importance of addressing protocol and hardware optimization in tandem.
We further note that (i) the slope of the linear fit for target (b) is larger than for target (a) and (ii) that the overall cost is also higher for target (b), implying that achieving high fidelities requires more improvement over the state of the art than achieving high rates.

SWAP-ASAP is used for all distances for target (a).
This is likely because meeting the more demanding rate target is challenging when employing purification.
Purification incurs a significant time penalty, as it (i) requires generating multiple entangled pairs and (ii) succeeds only probabilistically, with states having to be regenerated in case of failure.
Avoiding purification might accelerate the end-to-end entanglement generation process but results in lower-quality states.
For target (a) this is counteracted by employing the double-click protocol for all distances except $\SI{200}{km}$, as this protocol has no inherent limitations on the fidelity of the elementary link states generated, and can therefore enable higher fidelities than the single-click protocol.

Similar trade-offs can be observed for target (b).
The EPL protocol, which succeeds with a lower probability than DEJMPS, is employed at shorter distances.
However, at distances of $\SI{1200}{km}$ and above (which employ seven or more repeaters), DEJMPS is preferred.
This is because as more repeaters are used, more links must be purified, effectively resulting in more potential points of failure.
Therefore DEJMPS is a more appropriate protocol for this scenario, given that it typically succeeds with a higher probability than EPL.
A similar argument applies to the entanglement generation protocol change between $\SI{1200}{km}$ and $\SI{1600}{km}$ from double-click to single-click, respectively.
In other words, as the distance to be covered increases, one must use the protocols with the highest success probabilities i.e., DEJMPS instead of EPL and single-click instead of double-click.

We also note that the same number of repeaters is used for both targets at any given distance.
Furthermore, this number increases with the overall distance to cover.
Consequently, the internode distance stays approximately constant at around $\SI{100}{km}$ (see \cref{apx:waiting_time} for further details), which seems to indicate that this is the ideal spacing between repeaters in this particular scenario.
The exception is the case where the end-to-end distance is $\SI{200}{km}$, where no repeaters are used.

\subsection{Optimal hardware parameters}

\begin{figure*}
    \centering
    \includegraphics[width=\linewidth]{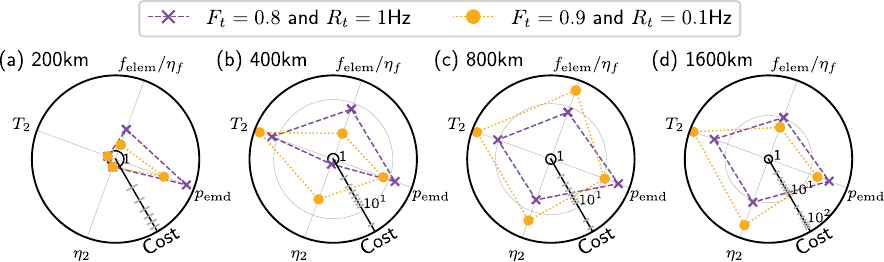}
    \caption{Parameter cost for the best solutions found at different total distances. The further from the center, the lines are towards a given parameter, the more improvement that parameter requires (logarithmic scale). The four parameters around the circle that we encounter in clockwise order: the elementary link fidelity $f_{elem}$ (\gls{dc}) or state efficiency $\eta_f$ (\gls{sc}), probability that an emitted photon is detected, given that it was not lost in fiber $p_{\text{emd}}$, two-qubit gate efficiency $\eta_2$, and coherence time $T_2$. The absolute value of the parameters can be found in \cref{tab:results_values} together with the protocols employed. Note the logarithmic scale.}
    \label{fig:paper_params}
\end{figure*}

\begin{table*}
    \centering
    \caption{Minimal hardware requirements and corresponding protocols per total distance. The elementary link fidelity is shown instead of the state efficiency $\eta_f$ for single click to allow for better comparison with double click. We also remark that all gate error $p_1$, $p_2$, $\xi_0$, $\xi_1$ and $p_{init}$ have been improved during the optimisation with the same cost shown in parenthesis. For simplicity, we have only included the final value of the two-qubit gate error. Finally, in the last column, we specify the protocols used as a combination of network protocol (SWAP-ASAP or BDCZ) plus entanglement purification (DEJMPS or EPL) if any plus entanglement generation (SC or DC). In the case of DEJMPS, we also specify the number of times a link is purified in the first nesting level with a newly generated link. We recall that entanglement purification is only applied to the first nesting level of the BDCZ protocol.}
    \label{tab:results_values}
    \begin{tabular}{r|lllllll}
        \multicolumn{8}{c}{(a) $F_t = 0.8$ and $R_t = \SI{1}{Hz}$} \\
        \toprule
        Distance & Repeaters & $f_{elem}$ ($\alpha$) & $p_{\text{emd}}$ (\%) & $p_2$ (\%) (Cost) & $T_1$ (h) & $T_2$ (s) & Protocols \\ \midrule
        200 & 0   & 0.8022 (0.16) & 39.55 & 2\% (1/5) & 1 & 1 & SWAP-ASAP + \gls{sc} \\
        400 & 3   & 0.9891 (--) & 66.09 & 2\% (1/5) & 1 & 12.78 & SWAP-ASAP + DC \\
        600 & 7   & 0.9810 (--) & 54.08 & 0.36\% (5.63) & 1 & 7.04 & SWAP-ASAP + DC \\
        800 & 7   & 0.9893 (--) & 77.51 & 0.35\% (5.751) & 1 & 10.47 & SWAP-ASAP + DC \\
        1200 & 15 & 0.9961 (--) & 69.24 & 0.58\% (3.47) & 1 & 8.41 & SWAP-ASAP + DC \\
        1600 & 15 & 0.9920 (--) & 85.28 & 0.037\% (57.28) & 1.47 & 22.84 & SWAP-ASAP + DC \\ \bottomrule
        \multicolumn{8}{c}{} \\
        \multicolumn{8}{c}{(b) $F_t = 0.9$ and $R_t = \SI{0.1}{Hz}$} \\ \toprule
        Distance & Repeaters & $f_{elem}$ ($\alpha$) & $p_{\text{emd}}$ (\%) & $p_2$ (\%) (Cost) & $T_1$ (h) & $T_2$ (s) & Protocols \\ \midrule
        200 & 0   & 0.9441 (0.034) & 16.68 & 2\% (1/5) & 1 & 1 & SWAP-ASAP + \gls{sc} \\
        400 & 3   & 0.7182 (0.26) & 49.56 & 0.41\% (3.22) & 1 & 9.68 & BDCZ + EPL + SC \\
        600 & 7   & 0.6795 (0.29) & 54.65 & 0.12\% (16.86) & 1 & 15.29 & BDCZ + EPL + SC \\
        800 & 7   & 0.9963 (--) & 60.99 & 0.13\% (15.68) & 1 & 28.84 & SWAP-ASAP + DC \\
        1200 & 15 & 0.9893 (--) & 78.58 & 0.12\% (19.98) & 1 & 97.17 & BDCZ + DEJMPS (1) + DC \\
        1600 & 15 & 0.7368 (0.074) & 60.67 & 0.04\% (50.50) & 1 & 112 & BDCZ + DEJMPS (2) + SC \\ \bottomrule
    \end{tabular}
\end{table*}
We now turn our attention to the improvement required for each hardware parameter. 
Concretely, we do so in \cref{fig:paper_params} for end-to-end distances between $\SI{200}{km}$ and $1600$ km. In each panel, we show in the radial axis the hardware cost of the hardware parameters needed to achieve the corresponding solution in \cref{fig:paper_everything}.
Starting from the top, the first hardware parameter shown is the elementary link fidelity $f_{elem}$ or state efficiency $\eta_f$ for DC and SC respectively.
We recall that the SC elementary link fidelity can be recovered using \cref{eq:paper_fsc} with the corresponding value of the bright-state population $\alpha$ being shown in \cref{tab:results_values}. Proceeding clockwise, the remaining parameters are the probability that an emitted photon is detected, given that it was not lost in fiber $p_{\text{emd}}$, the efficiency of two-qubit gates $\eta_2$, and the coherence time $T_2$. For simplicity, we have excluded the relaxation time $T_1$ from the figures as it is generally not improved beyond its state-of-the-art value (indicating that it is effectively already good enough). All parameter values can be found in~\cref{tab:results_values}.

As already seen in~\cref{sec:cost_scaling}, the best solutions for the lower-fidelity target (purple dashed line in~\cref{fig:paper_params}) do not make use of purification, as opposed to those for the higher-fidelity target (yellow dotted line in~\cref{fig:paper_params}). 
As a result, the best solutions for the lower-fidelity target have comparatively worse two-qubit gates and memory dephasing times.
This is to be expected, as the use of purification implies that (i) more gates will be executed and (ii) that states will spend a long time in memory as more pairs need to be generated.
In~\cref{tab:results_values} (a), we see an extreme example of this: the baseline gate quality is sufficient to span $\SI{400}{km}$ using SWAP-ASAP for $F_t = 0.8$ whereas two-qubit gates with error probability $\sim 0.4\%$ are required for $F_t = 0.9$ while employing EPL over the same distance.
On the other hand, the lower-fidelity-target solutions do require higher-quality elementary links and light-matter interfaces, to be able to (i) meet the more demanding rate target while also using \gls{dc} and (ii) achieve a relatively high fidelity without purification.

Overall, the probability that an emitted photon is detected, given that it was not lost in fiber, and $T_2$ are the parameters requiring the most improvement.
The first one is improved from a baseline value of $\sim 0.5\%$ to $> 50\%$ for all distances over $\SI{200}{km}$, and $T_2$ also requires 1-2 order-of-magnitude improvements over the $\SI{1}{s}$ baseline for all distances over $\SI{200}{km}$.
We note also that the hardware requirements for $\SI{200}{km}$ are significantly different than those for other distances.
This is to be expected, as for $\SI{200}{km}$ no repeaters are employed, which naturally results in significantly different hardware demands.
For example, memory quality becomes unimportant given that states are not stored for a significant amount of time, and gates do not require improvement either given that no operations must be performed.
In that case, only the probability that an emitted photon is detected, given that it was not lost in fiber requires significant improvement.

We must note that although we are only showing the two-qubit gate error in \cref{fig:paper_params}, all single-qubit gates and measurements have been improved by the same factors. This was done because even a perfect two-qubit gate was not sufficient to achieve $F_t = 0.8$ with $>15$ repeaters due to errors in other operations in the entanglement swap.

\section{Conclusion}
We have determined hardware requirements for chains of processing-node quantum repeaters spanning up to $\SI{1600}{km}$ while optimizing over protocols for entanglement generation, purification, and swapping strategy.
We have found that the overall hardware cost grows linearly with distance assuming a suitable choice of protocols is made.
This is surprising as various quantum-repeater-related quantities scale exponentially, such as photon loss in fiber and noise introduced in swapping.
A suitable protocol choice can then combat the exponential growth of hardware cost with distance that one would naively expect.
Such choices include, for example, employing purification to enable achieving higher-fidelity targets and a SWAP-ASAP strategy to attain higher rates.

The state-of-the-art parameters we considered were based on NV-center experiments.
However, the hardware model we used is abstract enough that the results we found are broadly applicable to any form of processing-node repeater.
Furthermore, the methodology we employed is fully general and can easily be adapted for use with different performance targets, hardware models, protocols, or network topologies.

In this work, we considered an idealized scenario in which all nodes are equally spaced with uniform fiber attenuation. However, real-world deployment of quantum networks will likely make use of existing fiber infrastructure~\cite{da2023requirements,rabbie2022designing}, for which this does not hold. In fact, it has been shown that taking the constraints imposed by real-world fiber networks into account significantly affects hardware requirements~\cite{da2023requirements}.
A natural extension would then be to apply our methods to such a network to investigate whether the linear scaling of hardware requirements we observed still holds.
Furthermore, one could also consider more intricate protocol choices. In real-world fiber networks, some links suffer from more attenuation than others. It might then be that it is wise to adopt a different purification strategy per link, as for example, it is more costly to regenerate a link in which attenuation is very high. Other examples of possible protocol choices that we have not explored are the use of cut-off timers and different swapping strategies. Allowing the genetic algorithm to also optimize over such protocols could enable further reductions in hardware requirements, although the associated growth of the parameter space could pose challenges. 

\section{Data Availability}
The data presented in this work have been made available at \url{https://doi.org/10.4121/0c6f100c-cf16-4fe1-81fe-afcafabdc7ca.v1}~\cite{2023replicationdata}.

\section{Code Availability}
The code that was used to perform the simulations and generate the plots in this paper has been made available at \url{https://gitlab.com/labay11/NetSquid-SimplifiedRepChain}~\cite{2023sourcecode}.

\section*{Acknowledgments}
This work was supported by the QIA-Phase 1 project which has received funding from the European Union’s Horizon Europe research and innovation programme under grant agreement No. 101102140. ALM is funded by the University of the Balearic Islands through the project BGRH-UIB-2021.

\onecolumngrid
\appendix
\section{Fidelity requirements for quantum key distribution}
\label{appendix:fidelity_qkd}
In this appendix, we show that states that meet the fidelity targets we set are also good enough to generate secret-key through the BB84 protocol. To do so, we will establish a connection between the maximum quantum-bit error rate (QBER) that can be tolerated in this protocol and the fidelity of the states. We assume that the states generated are depolarized Bell states, i.e.,
\begin{equation}
\rho =  W \ket{\phi^+}\bra{\phi^+} + \frac{1-W}{4}\mathds{I},
\end{equation}
where $W$ is related to the fidelity $F$ to the ideal Bell state $\ket{\phi^+} = \frac{1}{\sqrt{2}}\left(\ket{00} + \ket{11}\right)$ as $F = (1 + 3W)/4$ and $\mathds{I}$ is the four-dimensional identity matrix.
This is a simplification, as the states we generate are not of this form. However, any entangled state of fidelity $F$ to the ideal Bell state $\ket{\phi^+}\bra{\phi^+}$ can be brought to a state of this form with the same fidelity by application of random unitaries, a process known as twirling~\cite{horodecki1999general}. While one would never want to do this in reality, it is useful from a theory standpoint, as we can be sure that if a depolarized Bell state of fidelity $F$ is good enough to distill secret key, so is any other state of the same fidelity.

The QBER for a depolarized Bell state of parameter $W$ is given by $(1-W)/2$. This is due to the maximally-mixed component of the state, which has a weight of $1-W$. The probability of getting different outcomes when measuring two qubits in a maximally mixed state is $0.5$, regardless of the measurement basis, resulting in a QBER of $0.5$ for this state. Using the relation between the parameter $W$ and the QBER, and between $W$ and the fidelity $F$, we can derive the following relation between $F$ and the QBER, which we from here on out denote as $Q$:

\begin{equation}
    F = 1 - \frac{3Q}{2}.
\end{equation}

The secret-key rate SKR of the BB84 protocol is computed as (assuming that $Q$ is identical in both measurement bases, as is the case for depolarized Bell states)~\cite{bennett1992quantum}:
\begin{equation}
\text{SKR} = \text{R} \cdot \max\left\lbrace 0, 1 - 2 H(Q)\right\rbrace,
\end{equation}
where $R$ is the entanglement generation rate and  $H(p) = -p \log(p) - (1-p) \log(1-p)$ is the binary entropy function.

Using this expression, we find that the maximum QBER that still allows for a non-zero SKR is $\sim 0.11$, which for a depolarized Bell state corresponds to a fidelity of $\sim 0.84$. We further note that through two-way communication, the QBER threshold of the BB84 protocol can be raised to $0.2$~\cite{gottesman2003proof}, which corresponds to a depolarized Bell state fidelity of $0.7$.

Therefore, we conclude that states satisfying both of the fidelity targets we considered could also be used to distill secret key using the BB84 protocol.

\section{Waiting Time and maximum internode distance} \label{apx:waiting_time}
Here we compute the expected time required to distribute entanglement under different assumptions.
This is not strictly necessary as one can perform simulations to estimate the time required for entanglement distribution.
However, using such analytical results allows for saving computational resources.
For example, one might determine analytically that a target rate cannot be attained by a given repeater chain under the assumption that the only source of loss is attenuation in fiber.
In this case, performing the simulation is not necessary, as introducing more imperfections can only negatively affect the rate of entanglement generation.
    \paragraph{Without entanglement purification} The waiting time $T_{gen}$ is dominated by entanglement generation. This can be modeled with a geometric distribution, which determines the probability that the $t$-th attempt succeeds after $t - 1$ failed runs, \citep{brand2020efficient}
    \begin{equation} \label{eq:repchain_entgen_geom}
        P[T_{gen} = t] = p_{gen}(1 - p_{gen})^{t - 1}
    \end{equation}
    where $p_{gen}$ is the success probability of the entanglement generation protocol. Then, the average waiting time to create a link is
    \begin{equation} \label{eq:repchain_entgen_time}
        E[T_{gen}] = \frac{T_{cycle}^* + 2T_{com}}{p_{gen}}
    \end{equation}
    where $T_{com} = (L / 2) / c$ is the time needed for a photon to reach the detector placed equidistantly between two nodes separated by a distance $L$.
    For long distances, $T_{com} \gg T_{cycle}^*$ leads to an entanglement generation rate that decreases proportionally to $\exp(-L) / L$.
    
    \paragraph{With entanglement purification} To decrease the time taken by the purification process, we adopt a strategy that converts spatial resources to temporal resources. Specifically, at level $l$, instead of waiting for the generation of all $M$ links, we perform sequential purification as soon as two links are available. As a result, a link is purified $M$ times with newly created pairs that have suffered from less decoherence. Using this strategy, the expected waiting time $T^l$ at the $l$-th level needed to purify a pair $d = M - 1$ times can be calculated using the iteration formula \citep{dur_quantum_1999}
    \begin{equation}
        T^l_{k + 1} = \frac{T_k^l + T_0^l}{p_{suc}(\rho_k, \rho_0)}\qquad,\quad T_0^l = T^{l - 1},
    \end{equation}
    where $p_{suc}(\rho_k, \rho_0)$ is the probability of successful purification between a state $\rho_k$ that has been purified $k$ times and the new pair $\rho_0$. Here $1 / p_{succ}$ takes into account the number of repetitions until a single step succeeds to which we have to add the necessary time to create all previous pairs, $T_k^l$, as well as the new pair $T_0^l$.
    Concretely, for $l = 0$, $T_0^l$ corresponds to the entanglement generation waiting time in \cref{eq:repchain_entgen_time}. 
    
    Solving the recursion, one finds that the average waiting time to purify a pair $d$ times is
    \begin{equation} \label{eq:repchain_dist_time}
        E[T_d^l] = \sum_{j = 1}^d \left[ \prod_{k = j}^d T_0^l\frac{1 + \delta_{k1}}{p_{suc}(\rho_k, \rho_0)}\right] \approx \sum_{j = 1}^d \left[\frac{2T_0^l}{p_{suc}(\rho_0, \rho_0)}\right]^{d - j}
    \end{equation}
    where the approximation is valid when the success probabilities at the different $k$ steps are similar.

    \paragraph{Maximum internode distance}

    From the expected waiting time, it is possible to calculate the maximum distance for which the creation of an entangled state fulfilling the target rate and fidelity is possible.
    We remark that it is typically the rate target that cannot be met, due to the exponential decrease in the elementary link success probability, as well as the growth of the communication time with distance.
    It is possible to give an upper bound on this distance $D_{total}$ by considering the ideal case scenario of perfect quantum memory and photon detectors. Knowing the waiting time to create a single link \cref{eq:repchain_entgen_time}, the total waiting time to generate the end-to-end link with repeaters is at least $2 E[T_{gen}]$ because they can only perform one action at a time, as explained in \cref{sec:hardware_model}. Then, by finding the root of the function
    \begin{equation} \label{eq:results_levels_sc_root_r}
        R(L_{node}) = \frac{1}{2 E[T_{gen}(L_{node})]} - R_t = \frac{1}{2}\frac{\eta_{fiber}(L_{node}/2)}{T_{cycle}^* + L_{node}/c} - R_t\ ,
    \end{equation}
    where $L_{node} = D_{total} / (N_{node} - 1)$ and we assumed $\alpha = 0.5$, we can give an upper bound on the maximum distance.
    
    The above does not consider the target fidelity, as it depends on the strategy used, but for SWAP-ASAP we can give an upper bound using that
    \begin{align} 
        F_L = \frac{1}{4} + &\frac{3}{4}\left[\frac{(1-p_{1})^2 (1-p_{2}) (3 + 4(\xi_0\xi_1 - \xi_0 - \xi_1))}{3}\right]^{L - 1} \nonumber \left(\frac{4F - 1}{3}\right)^{L}\, \label{eq:qr_fid_n_swaps_perfect}
    \end{align}
    where $F_L$ is the fidelity after $L$ swaps and $F$ is the elementary-link fidelity.
    Hence, we can find the roots of the system of equations
    \begin{subequations} \label{eqs:results_levels_sc_root_fr}
        \begin{align}
        \frac{\alpha \eta_{fiber}(L_{node}/2)}{T_{cycle}^* + L_{node}/c} - R_t &= 0 \\
        \frac{1}{4} + \frac{3}{4} \left[\frac{4(1 - \alpha) - 1}{3}\right]^{N_{node} - 1} - F_t &= 0
    \end{align}
    \end{subequations}
    in terms of $\alpha$ and $D_{total}$.
    This method will give a much tighter bound because the trade-off between fidelity and rate is taken into account.
    Even tighter bounds on the waiting time have been studied \citep{brand2020efficient,coopmans2021improved}, but these simple cases are enough for our purposes.
    
    \begin{table}[ht!]
        \centering
        \caption{Upper bounds on the maximum end-to-end distance for which an entangled state meeting the target metrics can be generated assuming entanglement generation is done using the single-click protocol. Only photon loss is considered, assuming perfect quantum memories. The column on the left (rate only) corresponds to considering the rate bound only, whereas the one on the right takes the fidelity in the SWAP-ASAP case into consideration as well. The solutions correspond to the roots of \cref{eq:results_levels_sc_root_r} and \cref{eqs:results_levels_sc_root_fr}.}
        \label{tab:res_levels_max}
        \begin{tabular}{c|c|cc}
            \multicolumn{4}{c}{{\small \textbf{(a)} $F_t = 0.8$ and $R_t = \SI{1}{Hz}$}} \\ \toprule
            \textbf{Number of} & \textbf{Rate only} ($\alpha = 0.5$) & \multicolumn{2}{c}{\textbf{SWAP-ASAP}} \\
            \textbf{repeaters} & Distance (km) & Distance (km) & $\alpha$ \\ \midrule
            0 & 263 & 231 & 0.2 \\
            1 & 479 & 377 & 0.10774 \\
            3 & 958 & 669 & 0.05596 \\
            7 & 1917 & 1168 & 0.02852\\ \bottomrule
            \multicolumn{4}{c}{{\small \textbf{(b)} $F_t = 0.9$ and $R_t = \SI{0.1}{Hz}$}} \\ \toprule
            0 & 343 & 285 & 0.1 \\
            1 & 638 & 481 & 0.05179 \\
            3 & 1277 & 872 & 0.02636 \\
            7 & 2554 & 1563 & 0.01330 \\ \bottomrule
        \end{tabular}
    \end{table}
    
    The roots for the two methods are found using numerical methods, concretely the \texttt{hybr} algorithm implemented in the \texttt{scipy} python library. These are shown in \cref{tab:res_levels_max} for the two pairs of target values studied. Essentially, the first method gives the maximum possible distance while the second gives the maximum distance considering the trade-off typical of \gls{sc}, but only applicable to SWAP-ASAP, this second distance is always smaller than the former.
    
    One can note that the important parameter in the rate-only case is not the total distance but the internode length. In fact, dividing the distances in \cref{tab:res_levels_max}(a) by $N_{qr} + 1$ gives a maximum internode distance of $\SI{240}{km}$ and $\SI{319}{km}$ for the two pairs of targets, respectively. 

\section{Optimization Algorithm} \label{apx:optimization}
In this appendix, we go into more detail regarding the optimization method we employed and the choices we made.



Unfortunately, we cannot explore the whole space of parameters for computational reasons.
Therefore, we optimized only over the parameters that have a larger impact on the performance.
All the parameters over which we optimized can be seen in \cref{tab:res_gates_opt_params}.
We also wanted to compare hardware requirements with different numbers of repeaters for a given distance distance.
Thus, we performed different optimization runs for each distance, number of repeaters, and entanglement generation protocol.



\begin{table}
    \centering
    \caption{Parameters over which we optimized for (a) single- and (b) double-click entanglement generation protocols. The baseline value and the possible range of improvement are also shown for the hardware parameters. In terms of the protocol parameters, no baseline value is used and the range corresponds to the possible value it can take.}
    \label{tab:res_gates_opt_params}
    \begin{tabular}{c|cp{4.5cm}}
        \multicolumn{3}{c}{(a) Single click} \\
        \toprule
        Parameter & Baseline & Range \\ \midrule
        $\alpha$ & -- & $[\epsilon, 0.5]$ \\
        $\eta_f$ & $0.9196$ & $[0.9196, 1)$ \\
        $p_{\text{emd}}$ & $0.0046$ & $[0.0046, 1)$ \\
        $k_{gates}$ & 1 & $[1, 10^4]$ \\
        $T_1$ & $\SI{1}{h}$ & $[\SI{1}{h}, \SI{1e3}{h}]$ \\
        $T_2$ & $\SI{1}{s}$ & $[\SI{1}{s}, \SI{1e5}{s}]$ \\
        Strategy & -- &  SWAP-ASAP, \gls{epl}, DEJMPS ($n=1,2,3$ iterations) \\ \bottomrule
    \end{tabular}\hspace{1cm} 
    \begin{tabular}{c|cp{4.5cm}}
        \multicolumn{3}{c}{(b) Double click} \\
        \toprule
        Parameter & Baseline & Range \\ \midrule
        $f_{elem}$ & $0.92$ & $[0.92, 1)$ \\
        $p_{\text{emd}}$ & $0.0046$ & $[0.0046, 1)$ \\
        $k_{gates}$ & 1 & $[1, 10^4]$ \\
        $T_1$ & $\SI{1}{h}$ & $[\SI{1}{h}, \SI{1e3}{h}]$ \\
        $T_2$ & $\SI{1}{s}$ & $[\SI{1}{s}, \SI{1e5}{s}]$ \\
        Strategy & -- &  SWAP-ASAP, \gls{epl}, DEJMPS ($n=1,2,3$ iterations) \\ \bottomrule
        \multicolumn{3}{c}{}
    \end{tabular}
\end{table}%

The procedure for finding the optimal set of protocol and hardware parameters works as follows.
The \gls{ga} is initiated with a population of 120 individuals where each one of them consists of a vector $\vec{x}$ containing random values of the parameters in \cref{tab:res_gates_opt_params}.
Then, we use NetSquid to simulate the generation of an end-to-end link for each individual. From this simulation, we extract the fidelity of the final link $F$ and the time needed to generate it (with the entanglement generation rate $R$ being its inverse). However, since link generation is probabilistic, we do not use these values to compute the cost function. Instead, we repeat the simulation 200 times for $2$ and $3$ nodes and 100 times for a higher number of nodes and compute the mean fidelity $\bar{F}$ and mean rate $\bar{R}$.
These values were chosen for practical purposes as a balance between computational time and accuracy.
Hence, the combination of $\vec{x}$, $\bar{F}$ and $\bar{R}$ is used to evaluate the cost function in \cref{eq:cost_function}.
We note that the order is important because first calculating the cost and then averaging over all realizations might assign a very high-cost value to an optimal solution just because in one realization the performance targets were not reached.

We now move to the \gls{ga}.
This part of the process is divided into three stages that can be seen in \cref{fig:pp_repchain}.
First, we select the 24 individuals $\mathcal{S} = \{ \vec{x}_a \}$ with lower cost.
Second, 72 new individuals are created by crossing the parameters of two individuals randomly selected from among the 24.
This step splits two individuals $\vec{x}_a, \vec{x}_b \in \mathcal{S}$ in two parts $\vec{x}_{a,b}^{< k}$ and $\vec{x}_{a,b}^{\ge k}$ at a random position $k$ ranging from one to the number of parameters minus one.
Then, the new individual is created by combining the first part of $a$ and the second part of $b$, i.e. $\vec{x}' = (\vec{x}_{a}^{< k}, \vec{x}_{b}^{\ge k})$. Finally, the last 24 individuals missing to recover a population of 120 genes are created by choosing individuals from $\mathcal{S}$ and mutating a random parameter over a region close to the previous value.
The process is then repeated 500 times, which was found numerically to allow the convergence of all the situations studied.

After the final iteration, the individual $\vec{x}_{\mathrm{min}}$ with the lowest cost \eqref{eq:cost_function}.
However, there is no assurance of finding the global optimum, and there may be room for exploiting the minimum found. Therefore, we added an additional optimization step to attempt to further reduce the cost.
Concretely, we use a Hill climbing algorithm \cite{johnson1988easy} that searches for minimum solutions in a region around $\vec{x}_{\mathrm{min}}$.
This algorithm is a gradient-free method and it involves iteratively introducing minor modifications to a hardware parameter and assessing the resulting cost.
If the cost decreases, the modified parameter is retained, and the process is repeated by making further adjustments to the same parameter.
However, if the cost increases, the modification is discarded, and the algorithm proceeds to explore other parameters.
The extra minimization procedure allows us to exploit the minima found by the \gls{ga} assuming the optimal combination of protocols has already been found.
Hence, we only try to minimize the hardware parameters. The output of the local search algorithm is what we call in the paper the optimal solution.
We must note, however, that it is not possible to guarantee that the global optimum is found.

The simulations were executed on the High-Performance Computing facility in the Netherlands. The super-computer used is the Cartesius system which consists of nodes between 16 and 64 CPUs \cite{cartesius} in which we can parallelize the fitness evaluation of the individuals within a generation. The most common node contains $2\times 12$-core 2.4 GHz Intel Xeon E5-2695 v2 (Ivy Bridge) CPUs/node with 64 GB/node. For this reason, the population size will be a multiple of 12 (number of cores) to take advantage of the parallelization of the cost evaluation, concretely 120 individuals. The computational time increases exponentially with the number of nodes in the repeater chain.
A single execution of the simulation for 2 nodes takes $\SI{0.5}{ms}$, increasing to $\SI{100}{ms}$ for 9 nodes. Implying a total running time from $\SI{30}{min}$ for $2$ nodes to $\SI{3}{days}$ for $9$ nodes.


\section{Validation} \label{apx:validation}
In this work, we used an abstract model for repeater nodes with the goal of approximating the behavior of all types of processing nodes.
This makes results more broadly applicable and easier to interpret but comes at the cost of accuracy.
To get a sense for how well our abstract model performs, we compare results obtained using it with those obtained running a hardware-specific model simulating an NV-center repeater chain in~\citep{coopmans_netsquid_2020}.
There, the fidelity and entanglement generation rate was measured for a linear repeater chain with 0 (direct connection) and 3 repeaters. Entanglement generation was done using \gls{sc} and the network protocol used was SWAP-ASAP.
A star topology was considered, with the center qubit being optically active and used as a communication qubit.
All other qubits were used exclusively as memory qubits.
The different types of qubits have different coherence times~\cite{coopmans_netsquid_2020}.
Furthermore, induced dephasing noise was considered.
This was modeled by a dephasing channel which is applied to memory qubits whenever the communication qubit is used to attempt entanglement generation.

We, on the other hand, assume all qubits to be identical.
The properties of the qubits, namely coherence times and gate errors, were assumed to be given by the worst between memory and communication qubits in the NV case.
    
We assume the states generated have fidelity $(1 - \alpha)\eta_f$ to $\rho_{\text{sc}}$ (see Eq.~\ref{eq:single_click_state}), where we have condensed all the parameters that reduce the fidelity into $\eta_f = \frac{1 + \sqrt{V}}{2} (1 - p_{ph})$.
This accounts for visibility, the dephasing introduced due to double photonic excitation and phase uncertainty.
For comparison, we consider here also a simpler approximation, proposed in \citep{humphreys_deterministic_2018} which assumes perfect state efficiency, $\eta_f = 1$.
In both cases, we disregard dark counts and the bright-state population is set to $\alpha = 0.1$.
The three models are compared for two parameter sets.
The first consists of near-term hardware values, and the second of an improved set of parameters.
The exact parameters used are shown in \cref{tab:apx_nv_params_validation}. 
    
\begin{table*}
\centering
\caption{Parameters used in the validation plots shown in \cref{fig:results_validation}, same as those in \citep{coopmans_netsquid_2020}. In case there were differences in parameters between the memory and communication qubits, the most pessimistic value was chosen.}
\label{tab:apx_nv_params_validation}
    \begin{tabular}{@{}c|p{0.45\linewidth}|c|c@{}}
    \toprule
    \textbf{Parameter} & \textbf{NV equivalent} & \textbf{Near-term} & \textbf{Improved} $\mathbf{x10}$ \\ \midrule
     $p_{1,gate}$ & Carbon single qubit gate error & $(4/3)0.001$ & $(4/3)0.0001$ \\
     $p_{2,gate}$ & Electron-Carbon (EC) controlled $R_X$ gate error & $0.02$ & $0.002$  \\
     $\xi_0,\ \xi_1$ & Electron readout error & $0.05,\ 0.005$ & $0.005,\ 0.0005$ \\
     $p_{init}$ & Electron initialisation error & $0.02$ & $0.002$ \\
     $T_1$ & Electron relaxation time & $\SI{1}{h}$ & $\SI{10}{h}$ \\
     $T_2$ & Carbon dephasing time & $\SI{1}{s}$ & $\SI{10}{s}$ \\
     $T_{cycle}^*$ & Photon emission delay & \multicolumn{2}{c}{$\SI{3.8}{\mu s}$} \\
     $t_{1,gate}$ & Carbon single qubit gate duration & \multicolumn{2}{c}{$\SI{20}{\mu s}$} \\
     $t_{2,gate}$ & EC controlled $R_X$ gate duration & \multicolumn{2}{c}{$\SI{500}{\mu s}$} \\
     $t_{init}$ & Carbon initialisation duration & \multicolumn{2}{c}{$\SI{310}{\mu s}$} \\
     $t_{meas}$ & Electron read-out duration & \multicolumn{2}{c}{$\SI{3.7}{\mu s}$} \\
     $N_{qubit}$ & Number of qubits per node & \multicolumn{2}{c}{$4$} \\ \midrule
     $\gamma$ & Transmission loss & \multicolumn{2}{c}{$\SI{0.2}{dB/km}$} \\
     $c$ & Velocity of light in fiber ($n = 1.44$) & \multicolumn{2}{c}{$\SI{208189.207}{km/s}$} \\ \midrule
     $p_{\text{emd}}$ & Photonic efficiency excluding fiber & $0.0046$ & $0.58$ \\
     $V$ & Photon visibility & $0.9$ & $0.99$ \\
     $p_d$ & Probability of double excitation & $0.06$ & $0.003$ \\
     $p_\phi$ & Interferometric phase uncertainty & $\SI{0.35}{rad}$ & $\SI{0.11}{rad}$ \\ \bottomrule
    \end{tabular}
\end{table*}
    
\begin{figure*}[t]
    \centering
    \includegraphics[width=\linewidth]{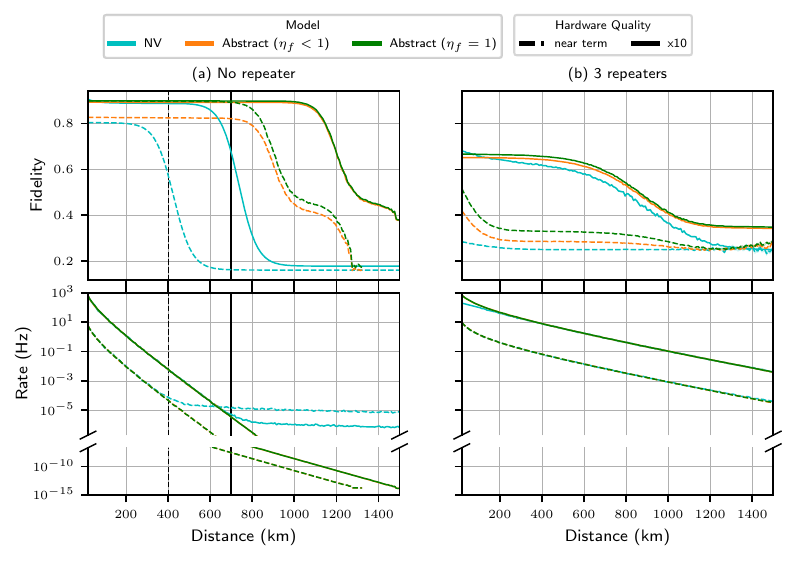}
    \caption{Comparison of NV-center model in \citep{coopmans_netsquid_2020} (cyan) with the abstract model using two approximations for the elementary link fidelity: in orange, with imperfect state efficiency; and in green, with perfect state efficiency. The distance is defined as the total distance between the end nodes. The vertical black lines denote the distance at which the transmission efficiency is comparable to the dark count probability. For 5 nodes, this occurs at a distance larger than the ones considered.}
    \label{fig:results_validation}
\end{figure*}

    
The fidelity and rate obtained with near-term and improved parameters for our abstract model and the hardware-specific NV model of \citep{coopmans_netsquid_2020} can be seen in \cref{fig:results_validation}.
We remark that the two abstract models only differ in the value of $\eta_f$, which only affects the elementary link fidelity, so no difference in rate is expected.
    
Starting with the no-repeater scenario, we see good agreement in the rate when $\eta_{trans} > p_{dc}$.
After this point, the rate in the abstract model continues to decrease exponentially, whereas the NV model stabilizes due to the presence of dark counts.
The most important difference occurs in the fidelity, even in this regime.
The NV model always gives a lower value, as expected, but for near-term hardware, the abstract model with perfect state efficiency deviates by more than $10\%$.
The more realistic model ($\eta_f < 1$) does give a better approximation when $\eta_{trans} \ll p_{dc}$, but we can see that the fidelity decreases significantly after $\eta_{trans} \approx p_{dc}$, showing that dark counts have a larger effect on the fidelity than on the rate.
Despite that, for distances where the achieved rate is higher than $\SI{0.1}{Hz}$, the difference between the expected fidelity is $< 3\%$ with imperfect state efficiency.
We note that this is the regime we investigate in this work.
With improved hardware, both abstract models give accurate values for the target metrics in the regime where dark counts can be neglected.

In the three-repeaters case, we see that the agreement in rate is good for all distances considered because the internode distance is never large enough for dark counts to become relevant.
On the contrary, similar to the previous scenario, the fidelity achieved is slightly higher than in the NV model, with the abstract model with $\eta_f < 1$ being the one that reaches a closer value. Nevertheless, the difference is much smaller than in the previous case due to the higher amount of operations and storage time needed, which take a much more important role than the errors introduced during the creation of elementary links.
However, there is one region with improved hardware where the fidelity in both abstract models falls below the NV one.
The same region shows the largest deviation from the expected rate. This can be due to the choice of parameters made, i.e., the fact that we considered the most pessimistic properties of electron and carbon qubits, but also the unrestricted topology.
At such short distances, the time spent mapping electrons to carbon states becomes important, resulting in a lower rate.
This is neglected in the abstract models, which means that fewer operations are performed there.
However, as we chose the worst parameters between memory and communication qubits when mapping from NV to abstract, more noise will be introduced in each operation.
Part of the disagreement in the fidelity can also be due to neglecting the induced dephasing in the carbon qubits.
Nevertheless, the agreement between the NV and the most elaborate abstract model is below $3\%$ when the rate is above $\SI{0.1}{Hz}$.

All in all, it is possible to conclude that the abstract model does give an accurate description of the rate in the regime $\eta_{trans} > p_{dc}$, showing that it is possible to disregard any restriction on the topology. The fidelity is better approximated with $\eta_f < 1$, although the difference between the two abstract models is reduced if three repeaters are used.

\twocolumngrid
\bibliography{biblio.bib}

\end{document}